\documentclass{article}
\usepackage[utf8]{inputenc}
\usepackage{arxiv}
\usepackage{multirow}
\usepackage{amsmath}
\usepackage{adjustbox}
\usepackage[T1]{fontenc}    
\usepackage{hyperref}       
\usepackage{url}            
\usepackage{booktabs}       
\usepackage{amsfonts}       
\usepackage{nicefrac}       
\usepackage{microtype}      
 \usepackage[numbers]{natbib}
\usepackage{graphicx}
\usepackage{float}

\title{Calculations of neutron fluxes and isotope conversion rates in a thorium-fuelled MYRRHA reactor, using GEANT4 and MCNPX}

\author{Asiya Rummana \\ 
University of Technology and Applied Sciences - Ibra, Oman\\
        E-mail: {\tt Asiya.Rumana@gmail.com}
        \and Roger John Barlow\\
         The University of Huddersfield, Huddersfield, UK\\
          E-mail: {\tt R.Barlow@hud.ac.uk}
           \and Syed Mohammad Saad\\
         University of Technology and Applied Sciences - Ibra, Oman\\
          E-mail: {\tt syedsaaad@ict.edu.om}
               }

\date{}
\begin{document}
\maketitle

\def \Iso #1 #2 {${}^{#1}{\rm #2}$}
\def \U {\Iso 233 U }
\def \Th {\Iso 232 Th }
\def \Pa {\Iso 233 Pa }
\def \keff {k_{\rm eff}}

\begin{abstract}
Neutronics calculations have been performed of the MYRRHA ADS Reactor with a thorium-based fuel mixture, using the simulation programs MCNPX~\cite{MCNPX} and Geant4~\cite{geant4}. Thorium is often considered for ADS systems, and this is the first evaluation of the possibilities for thorium based fuels using a reactor design which has been developed in detail. It also extends the application of the widely-used Geant4 program to the geometry of MYRRHA and to thorium.  An asymptotic \Th/\U\ mixture is considered, together with the standard MOX fuel and a possible \Th/MOX starter. Neutron fluxes and spectra are calculated at 
several regions in the core: fuel cells, IPS cells and the two (Mo and Ac) isotope production cells.
These are then used for simple calculations of the fuel evolution and of the potential for the incineration of minor actinide waste.
Results from the two programs agree and support each other and show that the thorium fuel is viable, and has good evolution/breeding properties, and that minor actinide incineration, though it will not take place on a significant scale, will be demonstrable.
\end{abstract}

\section {Introduction}
\subsection {Accelerator Driven Subcritical Reactors}

Accelerator Driven Subcritical Reactor (ADSR) systems have been much discussed since their original proposal by Bowman~\cite{bowman1992nuclear} and Rubbia~\cite{carminati1993report}. A core containing fissile isotopes is exposed to spallation neutrons from a high current particle beam (usually protons). $\keff$, the criticality of the core is below 1 -- typical designs have values in the range 0.90 to 0.98 -- which provides a multiplication factor $\frac{1}{1-\keff}$ (hence the name "energy amplifier") but never sends the system critical.  A full account is given in \cite{nifenecker2001basics} and \cite{nifenecker2003accelerator}. 

As well as their enhanced safety such systems also have the possibility to convert long-lived minor actinide (MA) waste to short-lived fission products, particularly when a fast neutron spectrum and the thorium fuel cycle are used, for four principal reasons.
\begin{itemize}
\item Neutron fluxes can be much higher than is possible in critical reactors: as discussed by Bowman~\cite{bowman1992nuclear} they can readily produce fluxes of $\sim 10^{16}{\rm n/cm}^2{\rm s}$, as opposed to $\sim 10^{14}$ for a power reactor or $\sim 10^{15}$ for a `high flux' reactor, as to the lack of heating from fissions means high flux can be attained at low thermal power, and safely below criticality.
    \item For fast neutrons, the ratio of fission cross sections to absorption cross sections is higher than in the classical thermal (water based) reactor situation.
    \item The production of further MA nuclei from \Th\ is less than for \Iso 238 U \ as more absorptions are needed and the cross sections are smaller.
    \item Variations in $\keff$ due to variations in reactivity do not have to be exactly compensated for.
\end{itemize}
Partitioning and transmutation in association with accelerator driven systems and in combination with geological disposal can provide a solution for the nuclear waste management problem~\cite{abderrahim2005fuel}.
ADSRs can burn not only their own minor actinides but also those produced by light-water reactors.

Many relevant experiments have been performed: the MUSE program at Cadarache,   GUINEVERE, Yalina at  Sosny, and  HYPER at  Seoul National University\cite{changizi2012neutronic}, culminating in the exposure of the CUCA critical assembly to a beam of protons, 
in which fission neutrons were observed\cite{ishi2010present}, albeit with a current of nanoamps rather than that milliamps that will be needed for a commercial ADSR system. 

\subsection{ADS systems and thorium}
ADSR with thorium fuel has gained large scale interest worldwide in the past two decades for energy production and waste transmutation, following the `Energy Amplifier' proposal of   Rubbia~\cite{carminati1993report}. Although ADSR are capable of burning any type of fuel, the choice of thorium provides the benefit of low radiotoxicity and proliferation resistance~\cite{THFC}. 

For the disposal of high activity nuclear waste, either the spent fuel can be sent for direct disposal (open cycle) or it can be reprocessed to extract transuranic and fission products (closed fuel cycle). The extracted species can then be transmuted into less radiotoxic or short-lived products. Comparisons of open fuel cycle, uranium-plutonium closed cycle and thorium-uranium closed cycle~   \cite{nifenecker2001basics} show that   closed fuel cycles produced significantly less radiotoxicity that open ones, and that  thorium-uranium cycle has  about two orders of magnitude lower radiotoxicity, for the first thousand years. 

The thorium fuel cycle   uses thorium as a fertile seed rather than \Iso 238 U .
Natural thorium (\Iso 232 Th )does not contain any fissile material ( unlike natural uranium which contains 0.7\% fissile \Iso 235 U) . It cannot be enriched in itself to produce materials of weapons grade so it poses a lower proliferation risk. Thorium can be combined with fissile isotopes ( \Iso 235 U \ or \Iso 239 Pu ) in nuclear reactors for conversion to the fissile \Iso 233 U .  
The analogue of the familiar uranium breeding chain which creates fissile \Iso 239 Pu

\hfil $^{238} {\rm U}\quad  
\begin{matrix} n  \\ \to \\  \\  \end{matrix} \quad 
^{239} {\rm U}\quad   
\begin{matrix} \beta  \\ \to \\  \\  \end{matrix} \quad  
^{239} {\rm Np} \quad   
\begin{matrix}  \beta \\ \to \\  \\  \end{matrix} \quad 
^{239} {\rm Pu}  
$\hfil

is

\hfil $^{232} {\rm Th}\quad  
\begin{matrix} n  \\ \to \\  \\  \end{matrix} \quad 
^{233} {\rm Th}\quad   
\begin{matrix} \beta  \\ \to \\  \\  \end{matrix} \quad  
^{233} {\rm Pa} \quad   
\begin{matrix}  \beta \\ \to \\  \\  \end{matrix} \quad 
^{233} {\rm U}  
$\hfil

which produces fissile \Iso 233 U . 
These chains are similar, the biggest difference being that the intermediate \Iso 233 Pa \ protactinium isotope has a half life of 27 days, much longer than the 2.4 days of the analogous  \Iso 239 Np .
 
\subsection{Minor Actinides}

The Minor Actinides (MA) are a major problem for radioactive waste. These are the transuranic elements neptunium (Np), americium (Am) and curium (Cm) which are generated by a combination of successive neutron capture and radioactive decays in a fission reactor. Although they are only few percent of the spent fuel, they are the most problematic part of the nuclear waste as they impose a long term environmental burden of their geological storage. They are highly radio toxic and their half - lives are up to millions of years. If the spent nuclear fuel is not reprocessed it must be treated as High level waste (HLW), and the cost and risk of storing this nuclear waste for a long time cannot be neglected~\cite{changizi2012neutronic}. 

Minor actinides can be destroyed by transmutation: the process of irradiating them in a high intensity neutron flux in order to decrease the long term radiotoxicity of the spent nuclear fuel. However the production of the neutron flux requires further fissions, and calculations are needed to decide whether the MA component destroyed is outweighed by the additional MA component  generated.

Transmutation can occur when the minor actinides are fissioned  with a single neutron interaction (direct) or through neutron capture(s) followed by fission (indirect). Since the reaction cross section for both direct and indirect fission tends to be very low, for  effective transmutation the neutron flux should be high and the irradiation time should be long~\cite{actinidetransmutation}.

Factors making a particular minor actinide a suitable candidate for  transmutation are:
\begin{enumerate}
  \item Long lifetime 
  \item High production level
  \item Neutron emission and decay heat in the final repository.
\end{enumerate}

Considering  the three elements in turn.

\textbf{Neptunium} (the predominant isotope is \Iso 237 Np ) is considered as secondary candidate for transmutation. It does not contribute to the decay heat output~\cite{actinidetransmutation}. \Iso 237 Np \ is a very long lived nucleide with a half - life of 2.144 million years. The production routes of \Iso 237 Np \ are:

a)	Two steps neutron capture on \Iso 235 U \ whose products are \Iso 236 U \ and \Iso 237 U .  \Iso 237 U \  (half – life = 6.75 days) finally decays to \Iso 237 Np . This reaction is preponderant in thermal reactors.

${}^{235}{\rm U} + n \to {}^{236}{\rm U}+ n \to {}^{237}{\rm U \to {}^{237}{\rm Np}}$

b)	90\% of neptunium production in fast reactors is through (n,2n) reactions on \Iso 238 U \ ~\cite{kooyman2018comparison}
which only happens above $\sim 6$ MeV, i.e. for fast neutrons.

${}^{238}{\rm U} + n \to {}^{237}{\rm Np} + 2n$

During irradiation process, neptunium either fissions or capture neutron to become \Iso 238 Np \  which is short lived with a half – life of 2.1 days. \Iso 238 Np \ decays to \Iso 238 Pu . The following reaction represents the transmutation steps of \Iso 237 Np :

${}^{237}{\rm Np} + n \to {}^{238}{\rm Np} \to {}^{238}{\rm Pu} $

\Iso 238 Pu \ being a strong alpha emitter is highly thermally active. Nevertheless, when mixed with existing plutonium it can be utilized as fuel since it is a neutron provider in a fast spectrum~\cite{kooyman2018comparison}.

\label{MATrans}

\textbf{Americium}: Due to its significant production level and gamma activity, Americium is considered as the prime candidate for transmutation. It has relatively short half - life and the dominant isotope in the irradiated nuclear fuel is \Iso 241 Am . Nevertheless, smaller but significant quantities of \Iso 242 Am \ and \Iso 243 Am \ are also produced. \Iso 241 Am \  is produced from the decay of \Iso 241 Pu \  which has a half-life of 14.4 years. \Iso 241 Am \  is consumed by absorption rather than fission, and the main product in the process is \Iso 238 Pu , from the successive decays of  \Iso 242 Am \ and  \Iso 242 Cm .  Further isotopes are also produced in small quantities ~\cite{kooyman2018comparison}. The transmutation of \Iso 241 Am \   involves the following reactions:

${}^{241}{\rm Am} + n \to {}^{242}{\rm Am}$

${}^{242}{\rm Am} \to {}^{242}{\rm Cm} \to {}^{238}{\rm Pu} \to {}^{239}{\rm Pu}$ (82.7\%)

${}^{242}{\rm Am} \to {}^{242}{\rm Pu}$ (17.3\%)

\textbf{Curium}: This is a major contributor to neutron emissions. It also significantly contributes to the gamma activity and radiotoxicity. However, its transmutation is generally ruled out due to low fission and capture cross sections of its principal isotopes, \Iso 242 Cm \ and \Iso 244 Cm  .

For these reasons, we consider \Iso 241 Am \ for transmutation studies that will be covered in section \ref{Am241Incineartion}.

\begin{figure}[ht]
\begin{center}
	\includegraphics[width=0.9\linewidth]{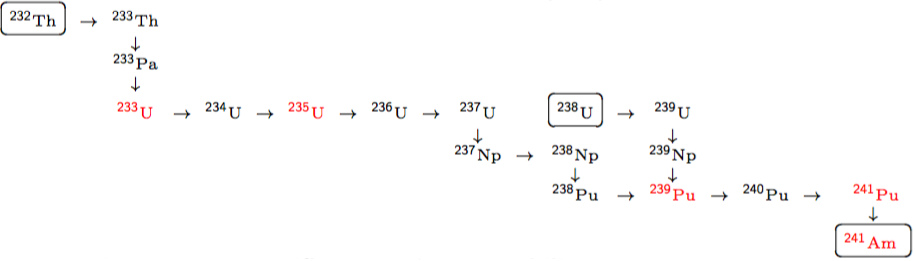}
	\caption{Minor actinides: principal pathways from \Iso 232 Th \ and \Iso 238 U . }
	\label{fig:MA Path}
\end{center}
\end{figure}

The thorium cycle produced less  MA waste, as can be seen from the pathways in Figure~\ref{fig:MA Path}. Absorption of a neutron moves one column to the right: beta decay moves one row down.  
Taking \Iso 241 Am \ as a typical MA isotope, the path starting at \Iso 238 U , as shown in Figure~\ref{fig:MA Path}, is much shorter (6 steps rather  than 14) than that from \Iso 233 Th .   Furthermore  the probability of neutron capture by \Iso 233 U \ is, at most energies,  about a factor of 10 less   than its fission probability, whereas for \Iso 239 Pu \ the comparable factor is smaller, in the range 2 to 3, as can be seen from the cross sections in the JEFF3.1N library~\cite{JEFF}.
 
 \subsection{Thorium ADSR studies}
 
Thus thorium is a promising alternative to uranium fuel in an ADSR because of its properties:proliferation resistance, abundance in nature and nuclear waste management.
 
Rather than designing a new thorium filled reactor for the present investigations, we used the design of reactor MYRRHA.   
The Belgian nuclear research Centre SCK$\cdot$CEN at Mol has proposed MYRRHA (Multipurpose hYbrid Research Reactor for High-tech Applications)~\cite{SCKCEN} which 
is currently under development and considerable detailed design work has been done, though this does not include studies using thorium fuel.
We have undertaken such simulations partly to increase our understanding of the possible uses of MYRRHA, and also to consider the problems of thorium fertile to fissile conversion and MA incineration using a mature and realistic ADSR design. 
 The studies were done using two simulation programs:  MCNPX and Geant4.
Early versions of these results have been previously published\cite{barlow2016studies,barlow2017simulations}.

\section {The MYRRHA Reactor} 

The MYRRHA reactor has been proposed
 to replace the Belgian Reactor 2 (BR2).
 It is a flexible design which can run either in critical or subcritical mode. It consists of an accelerator delivering a proton beam of 600 MeV in energy and about 4 mA beam current, and molten lead bismuth eutectic (LBE) as coolant which also acts as a spallation target and a subcritical core fueled with mixed oxide (MOX).  
  It is now approved and under construction and due deliver its
  first beams in 2026.

\begin{figure}[ht]
\begin{center}
\includegraphics[width=0.49\linewidth]{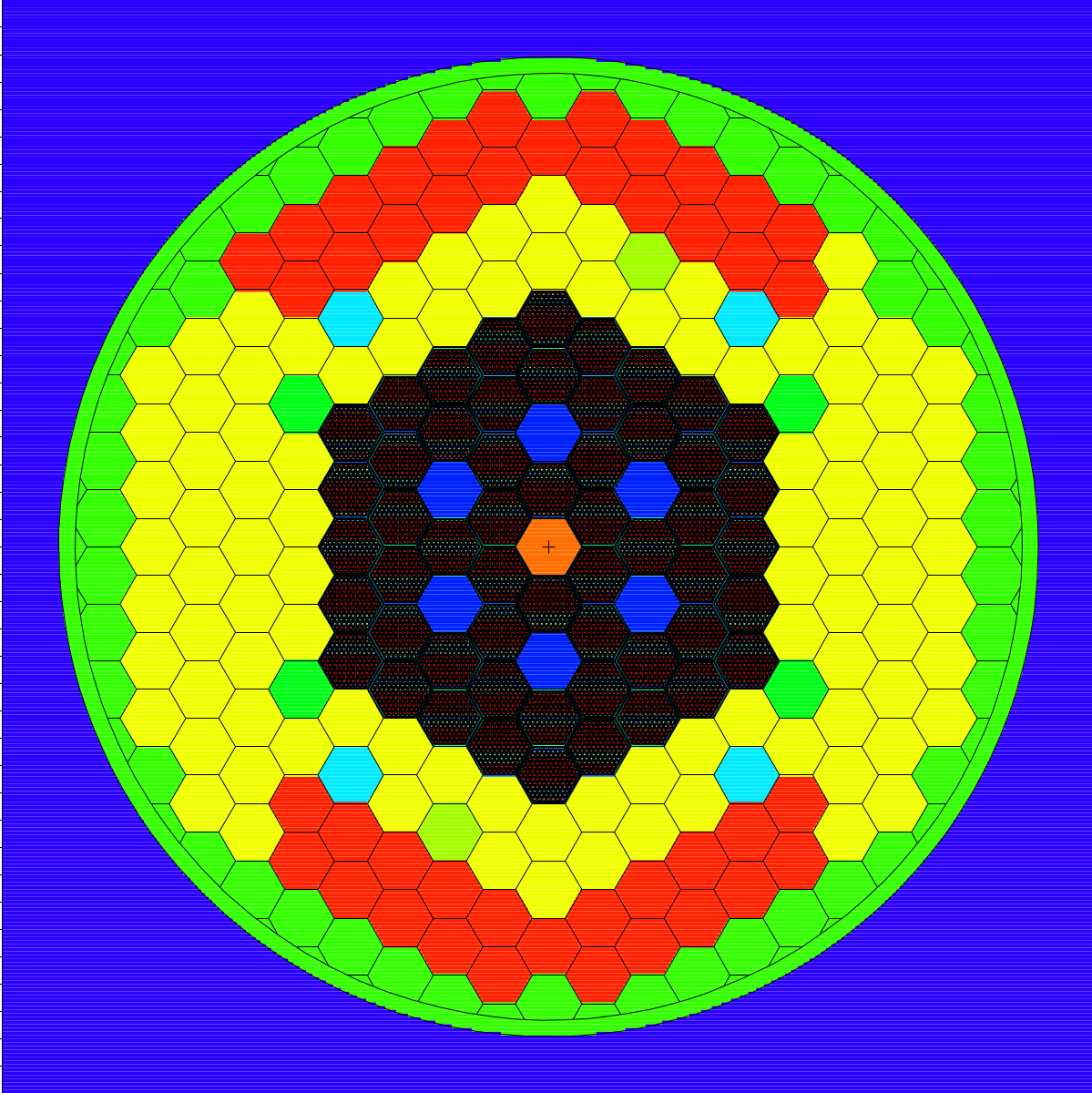}
\includegraphics[width=0.46\linewidth]{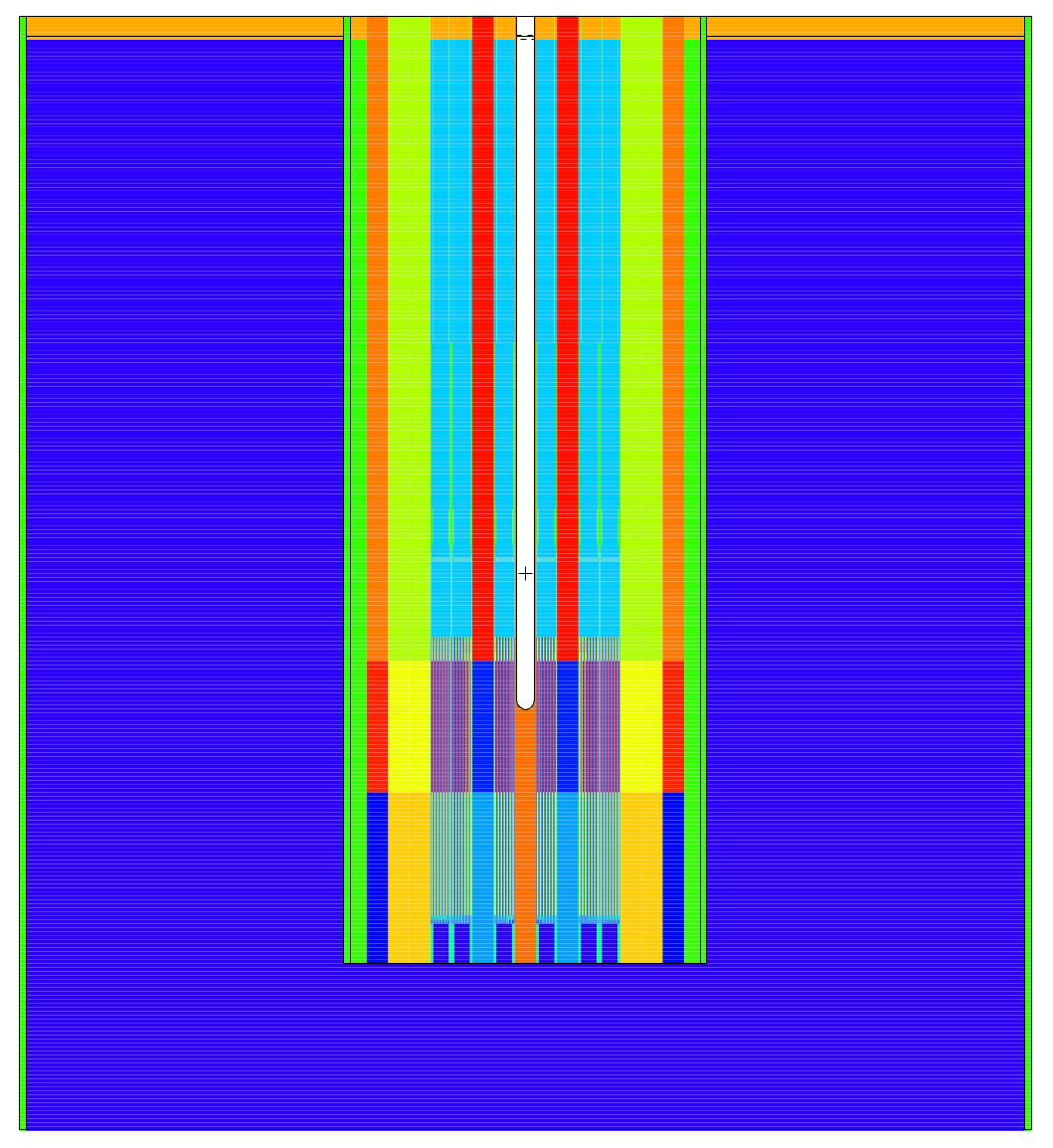}
\end{center}
\caption{MCNPX visualization of the full reactor considering a horizontal section (left) and vertical section (right). Colours (available in the online version) are explained in the text.}
\label{Fig:ReactorMX}
\end{figure}

 The geometry and material composition of the core is described by an MCNPX file  provided by the MYRRHA team \cite{MCNPXSpec} and shown in  Figure~\ref{Fig:ReactorMX}.
The cells shown in the above geometry are hexagonal cells with 10.45 cm between  opposite faces. Longitudinally the rods are divided into three parts, with the height of central active part being 65 cm. It is helpful to divide the cells into: 
\begin{enumerate}
  \item Inner cells, also known as core. Six fuel assemblies (black) surround the spallation target (orange). The next hexagonal ring contains 12 cells, half of which are fuel assemblies and half are In-Pile Section (IPS) cells (deep blue) for material testing in high neutron fluxes. Fuel assemblies in the next two rings make up the total 54 such cells. 
  \item Outer cells (yellow), which mostly contain lead bismuth eutectic (LBE). The next ring out is made up of 26 of these and  four cells (green) for control rods, not used in ADSR mode. The next ring is again LBE cells apart from  two cells (yellow-green) intended for the production of  Molybdenum isotopes and four cells (light blue) for Actinium. Rings after this also include Beryllium loaded reflectors (red) and finally stainless steel shielding (green).     
\end{enumerate}

\subsection{Fuel mixtures}

\label{sec:fuelmixtures}
\label{FuelMix}
The following three fuel mixtures were used in the simulations, with
 compositions  shown in Table~\ref{tab:CompositionMix}
:
\begin{enumerate}
  \item   U/Pu mixture (Mix 1): Standard MYRRHA MOX fuel  \cite{MCNPXSpec}. It consists of natural uranium plus plutonium (with a little americium) as obtained from fuel reprocessing. Oxygen is also included, as the fuel is used in the oxide form rather than being purely metallic.
  \item Th/Pu mixture (Mix 2): All the uranium in Mix 1 is replaced by thorium, which naturally occurs as pure \Iso 232 Th. This 
  represents a possible initial fuel mixture before any \Th \ to \U \ conversion has occurred.
  \item 	Th/U mixture (Mix 3): All the Pu and Am in Mix 2 is replaced by \Iso 233 U . This represents an asymptotic fuel mix in which the initial MOX starter has been consumed, but has been replaced by \U\   produced from the \Th .  
\end{enumerate}

\begin{table}[ht]
\begin{center}
\begin{tabular}{|c|c|c|c|}
\hline
Fuel mix 
& Element & Percentage  & Percentage\\
& & MCNPX & Geant4 \\
  \hline
Mix 1 & \Iso 16 O  & 11.6718  &\\
& \Iso 234 U  & 0.0033 &\\
& \Iso 235 U  & 0.4395 &\\
& \Iso 238 U  & 61.3717 &\\
& \Iso 238 Pu  & 0.6083 & Same composition \\
& \Iso 239 Pu & 14.8343 &\\
& \Iso 240 Pu  & 7.0417 &\\
& \Iso 241 Pu  & 1.5924 &\\
& \Iso 242 Pu  & 2.0066 &\\
& \Iso 241 Am  & 0.4304 &\\
 \hline
Mix 2 & \Iso 16 O & 11.6718 & 11.6718  \\
& \Iso 232 Th  & 58.4071 & 62.0000 \\
& \Iso 238 Pu  & 0.6865 & 0.6083 \\
& \Iso 239 Pu  & 16.7407 & 14.8343 \\
& \Iso 240 Pu  & 7.9467 & 7.0417 \\
& \Iso 241 Pu  & 1.7970 & 1.5924 \\
& \Iso 242 Pu  & 2.2645 & 2.0066 \\
& \Iso 241 Am  & 0.4857 & 0.4304 \\
 \hline
Mix 3 & \Iso 16 O & 11.6718 & 11.6718  \\
& \Iso 232 Th  & 70.1260 & 59.8200 \\
& \Iso 233 U  & 18.2022 & 16.0000 \\
 \hline
\end{tabular}
\caption{Compositions (percentages, by number) for the  three different fuel mixes considered }
\label{tab:CompositionMix}
\end{center}
\end{table}

\def \tkeff {$k_{\rm eff}$}

Mix 1, the standard MYRRHA fuel mix, has a \tkeff\  of $0.95178 \pm 0.00059$ as evaluated by the MCNPX {\tt KCODE} process\cite{MCNPX} (using 1000 cycles of 1000 particles, ignoring the initial 100 cycles).  For mix 2, simply replacing all the uranium by thorium gives a very different \tkeff, so the relative fractions of thorium and the fissile Pu/Am mix were adjusted
to give a value approximately 0.95 - actually $0.94167 \pm 0.00061$. This was used for the Geant4 studies.   
For the MCNPX studies a further slight adjustment brought \tkeff\ to $0.95165  \pm 0.00058$,
much closer to the equivalent of mix 1.

A similar adjustment was made for mix3, the mixture used in Geant4 giving $\keff=0.96235 \pm 0.00061$ and the MCNPX mixture $0.95096 \pm 0.00068$.


\section{Simulation programs}

\subsection{MCNPX}

MCNPX~\cite{MCNPX} is a well established program widely used in reactor simulations. Version 2.7.0 was used, with the standard ENDF/B-VII cross section libraries \cite{endf70}. The geometry was 
specified by a input deck supplied to us by Edouard Malambu and Alexey Stankovsky of SCK$\cdot$ CEN, dated August 2014~\cite{MCNPXSpec}
and taken from drawings and specifications version REV.1.6.
The geometry shown in 
Figure~\ref{Fig:ReactorMX}
uses this deck.
Various different configurations have been proposed in the past, but the configuration is now stable. 
Note in particular that the target occupies only a single cell, whereas in earlier designs it occupied several.

Each simulation was done using 10,000 initial protons and took typically 12 hours to run. 

\subsection {Geant4}

Geant4~\cite{geant4} is a program originally used to simulate particle physics detectors, but which has since been extended to many different fields. We used version 
4.10.1, the latest release (December 2014) at the start of the study. 
The 
  physics list QGSP$\_$BIC$\_$HP was used,  based on earlier studies (~\cite{barlow2015simulations},~\cite{barlow2016characterisation},~\cite{rummana2016simulation}). 
  100,000 beam protons were used for the simulations. These protons are sufficient to provide statistically reliable results that can be compared with the results predicted by earlier studies \cite{sarotto2013myrrha,barlow2016studies,barlow2017simulations}. 

\subsubsection{Cross sections for transuranic elements}\label{Transuranic}

Because of its history, 
GEANT4 did not handle
 transuranic elements: the standard G4NDL library  does not provide data for isotopes having atomic number Z\textgreater 92.
 However for reactor studies, such as this one, these cross sections are needed.
 
 We therefore used the program of Mendoza et al \cite{mendoza2012new,mendoza2014new}to transform the widely used JEFF 3.1 library\cite{JEFF}, which uses neutron cross sections from ENDF/B-VII  and contains the relevant transuranic cross sections,
  into G4NDL format.
  Three environmental variables had to be specifically set:
  
a)	{\tt G4NEUTRONHP$\_$SKIP$\_$MISSING$\_$ISOTOPES=1} 

b)	{\tt G4NEUTRONHP$\_$DO$\_$NOT$\_$ADJUST$\_$FINAL$\_$STATE=1} 

c)	{\tt AllowForHeavyElements=1}. 

Even so there were problems: the program would run for several events and then crash with the message 

{\tt Called G4PiNuclearCrossSection outside parametrization}

		or 

		{\tt ***G4ElectroNuclearCrossSection::GetFunctions: A=``$<<$244.064$<<$"(?). No CS returned!}

Investigation showed that these messages were generated from code relating to the calculation of
cross section formulae for pions and  muons,
which could not handle isotope nuclei with $Z>92$. 
The 600 MeV beam energy is above the 289 MeV threshold energy for pion production, so pions and consequently muons are generated in our simulation, whereas others who have used Geant4 for conventional reactor studies where the energies do not exceed a few MeV will not have encountered this problem.

The relevant source code files were found to be 
{\tt G4KokoulinMuonNuclearXS.cc}, {\tt G4KokoulinMuonNuclearXS.hh} and {\tt G4PiNuclearCrossSection.cc}
and these were modified, after
discussion with the Geant authors
~\cite{priv}  to use $Z=92$ in their formulae for nuclei with $Z>92$.

 While this is an approximation, it is a small effect that applies to a small number of targets and a very small number of particles.
 In all our fuel mixtures the fraction of transuranics is only a few percent, and the electromagnetic  interaction of a pion with, say, one of the few americium nuclei is not going to be that different to its interaction with uranium. 
 At 600 MeV and below the numbers of  pions and muons are very small.  An MCNPX study showed that for 10000 beam protons, 9,830,554 neutrons were produced, but only 438 charged pions are produced.  A slight approximation to the behaviour of these 438 pions/muons is not going to affect the evaluated flux and energy spectrum of the neutrons. The overall effect on the results will be negligible.

There was also a warning produced - ``\#\#\# G4SeltzerBergerModel Warning: Majoranta exceeded! ", but it does not cause the program to crash and investigation showed it was not serious.

\subsection {Implementation of the MYRRHA geometry in GEANT4}

To implement the reactor geometry in GEANT4, a bottom-up approach was followed in which the fuel pin was considered as the smallest unit: pins are constructed with their gaps and cladding.

In the next step, fuel assemblies are created by arranging the pins in a hexagonal lattice. Different types of cells are also created such as the  In Pile Section (IPS) cells and the spallation target. 

These cells are then arranged in another lattice to build up the whole core. After the  core, 5 outer rings are constructed including Lead bismuth eutectic (LBE) cells, Mo/Ac cells, control rods, reflectors and steel shielding cells.

GEANT4 has a variety of visualization drivers offering specific features to meet different demands of its users. In the present simulations, the OpenGL driver is used due to its fast visualization feature for demonstrating geometries, trajectories and hits \cite{DetSimVis,Allison2013Geant4}. 
OpenGL offers interactive features: zoom, rotate and translate. However, for the complicated dimensions (Radial dimensions of a few millimeters and vertical dimensions of many centimeters) of the reactor, `zoom' did not prove useful. Instead images were displayed with the radial dimensions enlarged temporarily  for the purpose of better visualization. Such displays are shown below, and this
system proved to be helpful in designing and debugging~\cite{asiya2019}. 

\subsubsection{The Fuel pin}
\begin{figure}[ht]
\begin{center}
\includegraphics[width=0.59\linewidth,clip,trim={100 50 30 170} ]{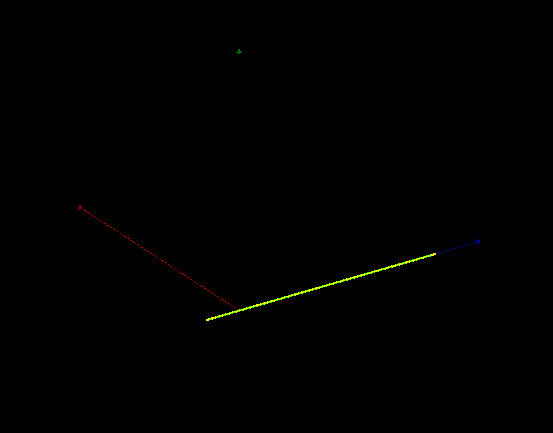}\ \includegraphics[width=0.3\linewidth,clip,trim={0 50 0 50}] {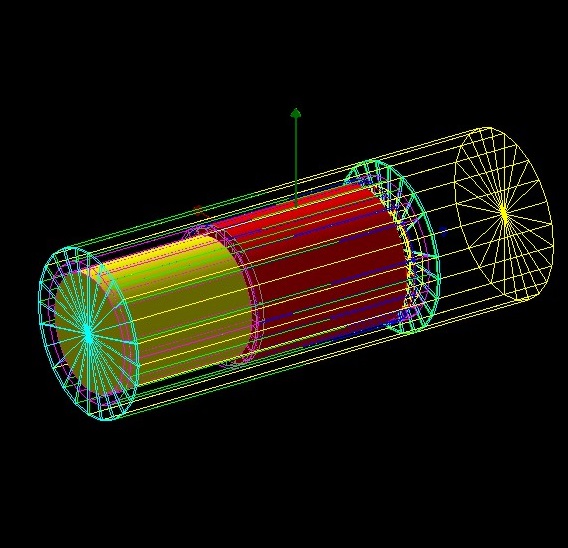}
\end{center}
\caption{GEANT4 visualization of fuel pin: original dimensions (left) and enlarged dimensionas (right)} 
\label{Fig:Pin1}
\end{figure}

The fuel pin is  a rod with cylindrical cladding loaded with cylindrical fuel pellets. Figure~\ref{Fig:Pin1} shows the implementation of the fuel pin in GEANT4, with the original dimensions and in the enlarged version, for which the detail can be  seen: it consists of three regions with 65 cm as length of the central active region, in which fuel (as described in section \ref{FuelMix}) is loaded.
Dimensions are taken from 
the specifications provided by \cite{MCNPXSpec}. The structure of the fuel pellets is not simulated, the fuel is taken as being continuous in the rod. 
The upper and lower regions include insulator segments and a gas plenum chamber.  

\subsubsection{Fuel assembly}

 \begin{figure}[ht]
\begin{center}
\includegraphics[width=0.8\linewidth]{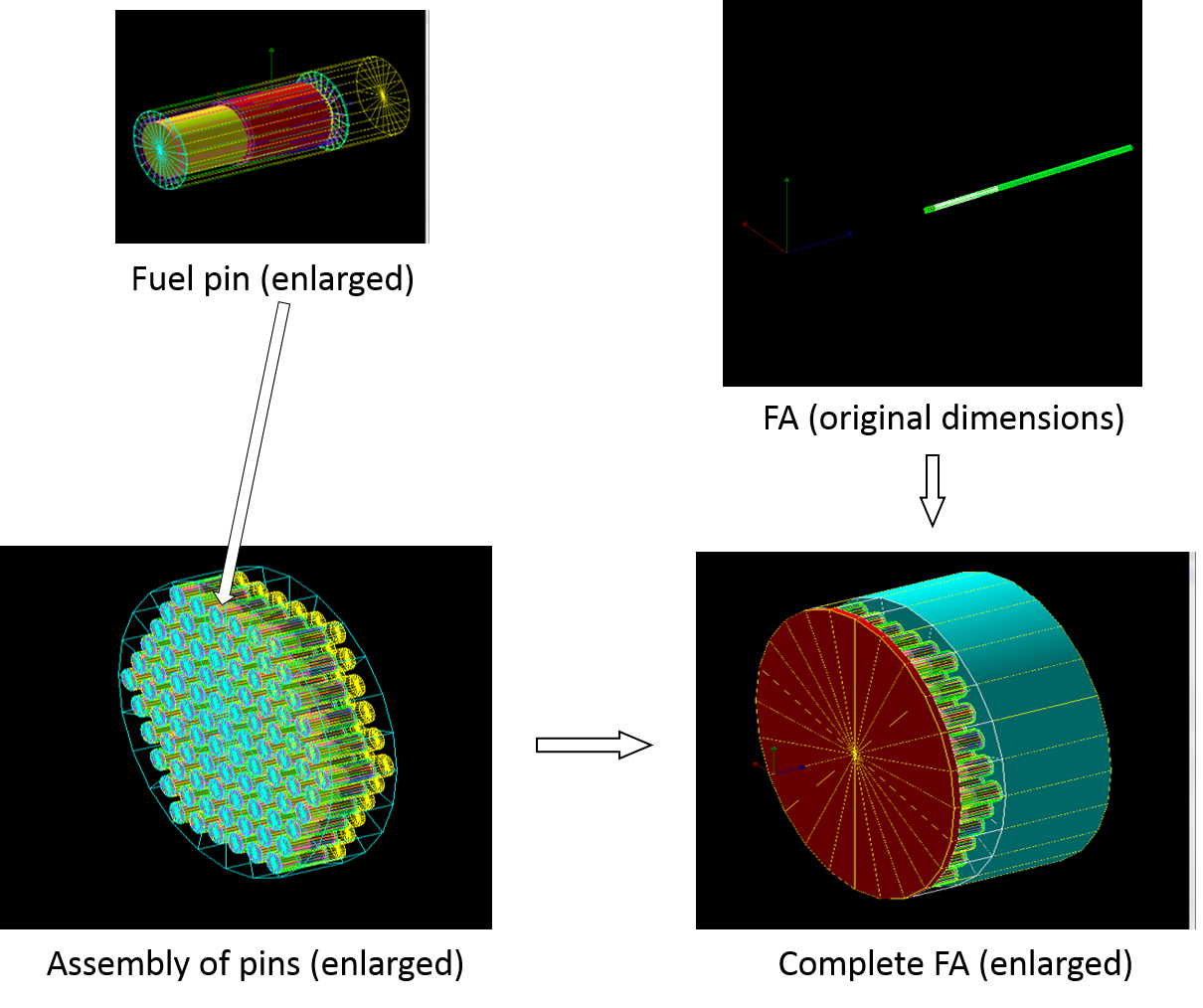}
\end{center}
\caption{GEANT4 visualization of fuel assembly (FA) }
\label{Fig:FA}
\end{figure} 

Each fuel assembly is the collection of 91 fuel pins that are arranged in a hexagonal bundle. Figure~\ref{Fig:FA}  shows the steps in the creation of the fuel assembly from the fuel pin. In the enlarged picture (Complete FA), the blue and red cylindrical shape shows the upper and lower part of the fuel assembly respectively and the central region is the active part of the fuel assembly.

\subsubsection{IPS cells}
\begin{figure}[ht]
\begin{center}

\includegraphics[width=0.4\linewidth]{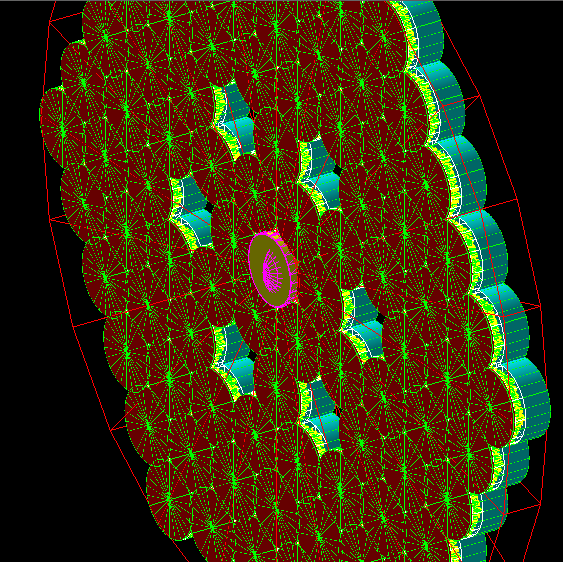}
\ \ 
\includegraphics[width=0.4\linewidth]{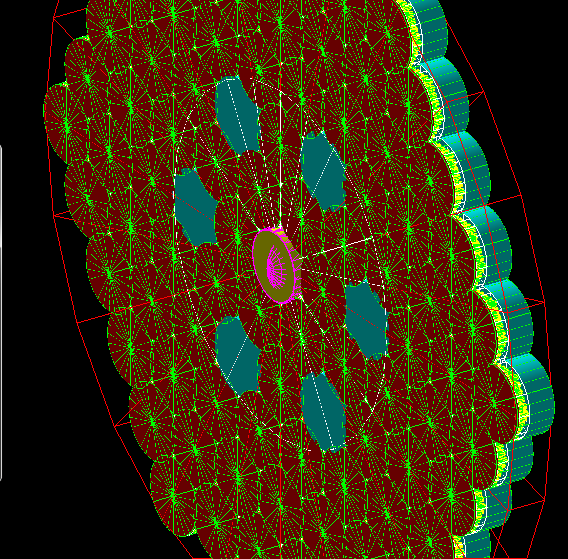} 

\caption{IPS assemblies:  (left) empty locations for IPS (enlarged dimensions)and (right)  IPS placed in the core } 
\label{Fig:IPS1}
\end{center}
\end{figure}

Six IPS cells
are created and placed in the reactor core as shown in the Figure~\ref{Fig:IPS1}. Initially in the  program the reactor core is created with all the fuel assemblies, then the designated locations are made empty   
and IPS cells are placed at these locations.
\subsubsection{Central cell - the Spallation target}
\begin{figure}[ht]
\begin{center}
\includegraphics[width=0.45\linewidth]{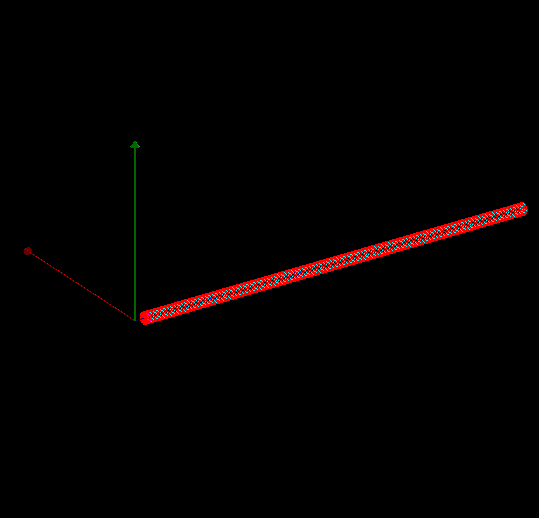}
\  \ 
\includegraphics[width=0.45\linewidth]{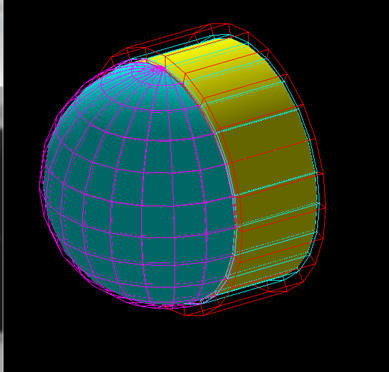}
\end{center}
\caption{GEANT4 visualization of spallation target: Original (left) and Enlarged (right) dimensions} 
\label{Fig:Target2}
\end{figure}

The Spallation target is one of the key component of the ADSR. The beam tube ends on hemispheric window as shown in  Figure~\ref{Fig:Target2}. The target is constructed using {\tt G4Tub} shape for the beam tube and {\tt G4Sphere} for the hemispherical end, with appropriate coordinates for the adjacent placement of these two sections of the target.  The spallation target cell is then placed with the hemispheric end located at the center of the core.    

\subsubsection{The complete reactor}
\begin{figure}[ht]
\begin{center}
\includegraphics[width=0.45\linewidth]{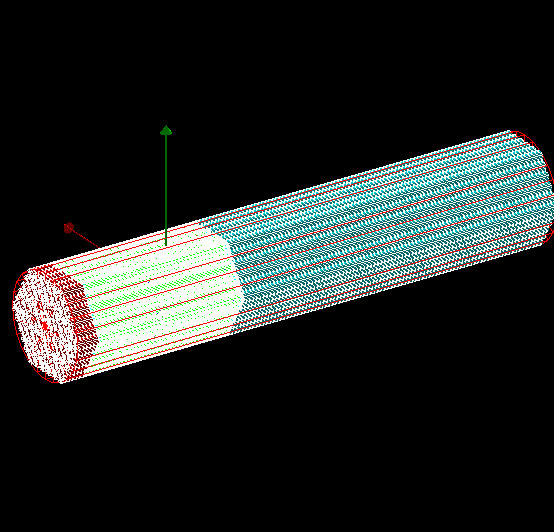}
\ \ 
\includegraphics[width=0.45\linewidth]{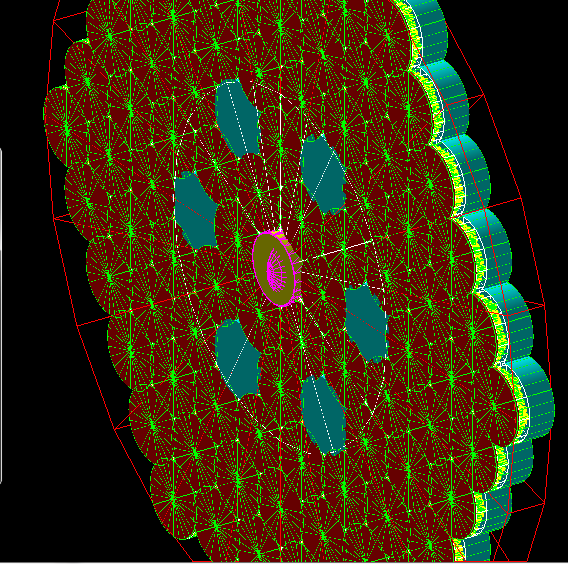}
\end{center}
\caption{GEANT4 visualization of reactor core: Original dimensions (left) and enlarged dimensions (right)} 
\label{Fig:Core1}
\end{figure}

The inner cells include total 54 fuel assemblies and six IPS cells constituting the reactor core, as shown in Figure~\ref{Fig:Core1}.

The outer cells are then added. These have the same dimensions but are filled with different materials for LBE cells, Mo-Ac cells, Beryllium reflectors and Stainless Steel shielding.

\begin{figure}[ht]
\begin{center}
\includegraphics[width=0.45\linewidth]{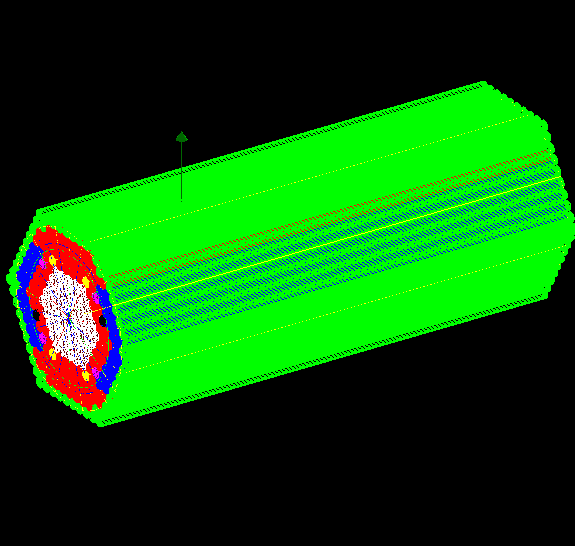}
\ \ 
\includegraphics[width=0.45\linewidth]{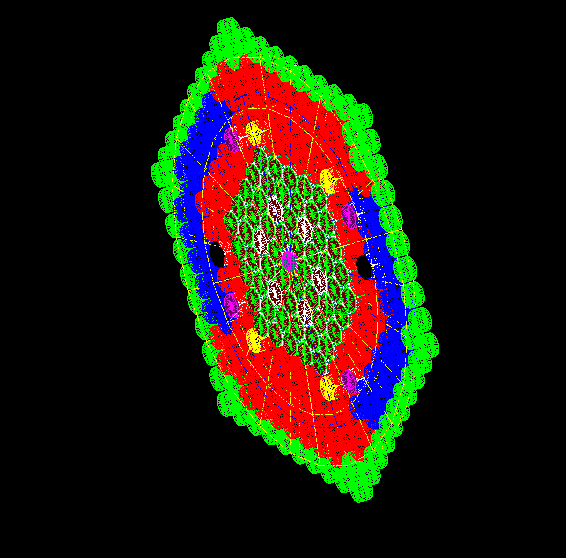}
\end{center}
\caption{GEANT4 visualization of the reactor: (left) Original dimensions and (right) Enlarged Dimensions} 
\label{Fig:Reactor1G4}
\end{figure}

The complete picture of the reactor as modelled in GEANT4 with all the cells placed, is presented in Figure~\ref{Fig:Reactor1G4}.  
It should be compared with the MCNPX model of Figure~\ref{Fig:ReactorMX}

 \section {Results: Neutron Fluxes and Energy Spectra}

\subsection{Neutron Flux}
Average neutron fluxes were evaluated for the fuel cells, as this is important for fuel evolution, and for the three types of cell intended for irradiation: the IPS cells in the core and those for Molybdenum and Actinium production in the outer region. Neutrons of energy range $10^{-8}$ MeV to 100 MeV were considered.
MCNPX results are obtained from the F4 neutron flux tally; for Geant4, the function {\tt  G4PSPassageCellFlux} is used to record the flux.

Values are shown in 
Table~\ref{tab:Flux} for the 3 different fuel mixtures.
Numbers shown are normalised to a nominal 1 mA proton beam current.
Statistical errors are of order 5\%. (The errors are calculated and printed for MCNPX: for Geant4  they were estimated by running the  same job twice with a different random seed and examining the differences.)

The table shows that GEANT4 and MCNPX results are in fair agreement. There are differences, but there is no discernible general trend. It is not possible to 
assign causes to these differences - whether they are due to the details of the programs used, or the data libraries, or the slight difference in the value of $k$\textsubscript{eff}, as described in Section~\ref{sec:fuelmixtures}, for Mixtures 2 and 3. 

The simulations show similar fluxes for the inner IPS cells and the fuel cell average, with lower values for the Mo cells and even lower ones for the Ac cells further out. There are differences between the different fuel mixtures but no general trend can be discerned. 

\begin{table}[ht]
\begin{center}
\begin{tabular}{|c|c|c|c|c|c|c|c|}
\hline
Location 
 & \multicolumn{2}{c|}{Mix1} & \multicolumn{2}{c|}{Mix2} & \multicolumn{2}{c|}{Mix3} \\
 \cline{2-7} 
& G4 & MX & G4 & MX & G4 & MX \\
 \hline 
Fuel & 6.28 & 7.08 & 6.37 & 5.80 & 8.62 & 8.41\\
 \hline
IPS & 5.19 & 8.48 & 8.76 & 6.98 & 8.34 & 9.95\\
 \hline
Mo cell & 2.8 &4.8 & 5.07 & 3.91 & 6.94 & 6.04 \\
 \hline
Ac cell & 1.17 & 1.09 & 0.93 & 0.90 & 1.98 & 1.37 \\
 \hline
\end{tabular}
\caption{Average flux values (units are $10^{14}$ neutrons/cm\textsuperscript{2}/s) for a 1 mA proton beam}
\label{tab:Flux}
\end{center}
\end{table}

\subsection{Neutron energy spectra}\label{Spectra}

\begin{figure}[ht]
\begin{center}
\includegraphics[width=0.49\textwidth]{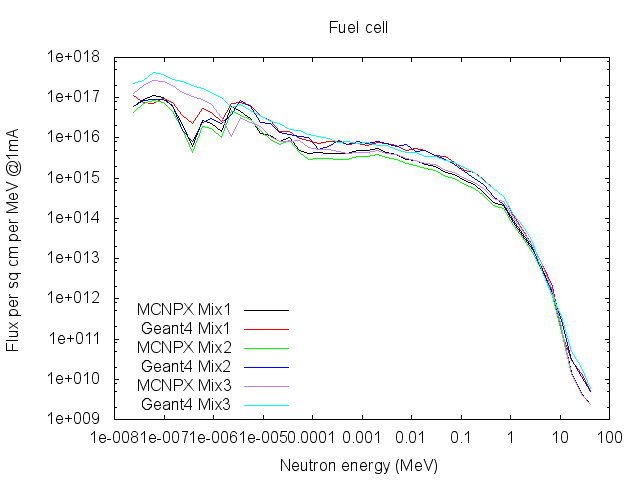}
\includegraphics[width=0.49\textwidth]{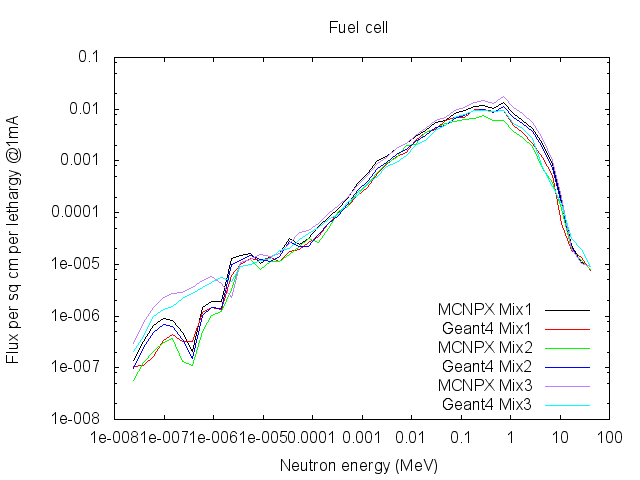}
\end{center}
\caption{Neutron flux averaged over the fuel cells per unit energy(left) and lethargy (right)}
\label{Fig:Fuel}
\end{figure}

We considered the energy spectra 
for the three fuel mixes as obtained by GEANT4 and MCNPX.
The energy spectra for the fuel is presented in Figure~\ref{Fig:Fuel} in terms of both energy and lethargy\footnote{Lethargy is  defined as the logarithm of the ratio of maximum energy that a neutron might have in a reactor to the neutron energy}.  
The fuel cell spectra show a hard component – all the way up to the proton energy -  though energies go all the way down to thermal energy. The spectra and the overall numbers agree well for GEANT4 and MCNPX. This is true for all the three fuel mixtures. Nevertheless they differ in detail:
the most marked regions of disagreement being at low energies in the Fuel cells and at high energies in the IPS cells. Whether this is due to differences in the algorithms or in the data libraries is not clear, but  detailed predictions involving these neutrons should be treated with appropriate caution. The fluctuations in the spectrum do not stem from low simulation statistics rather due to the energy dependence of the cross section.  
The results for the standard mixture are compatible with the earlier studies of Sarotto et al\cite{sarotto2013myrrha}
 
\begin{figure}[ht]
\begin{center}
\includegraphics[width=.49\textwidth]{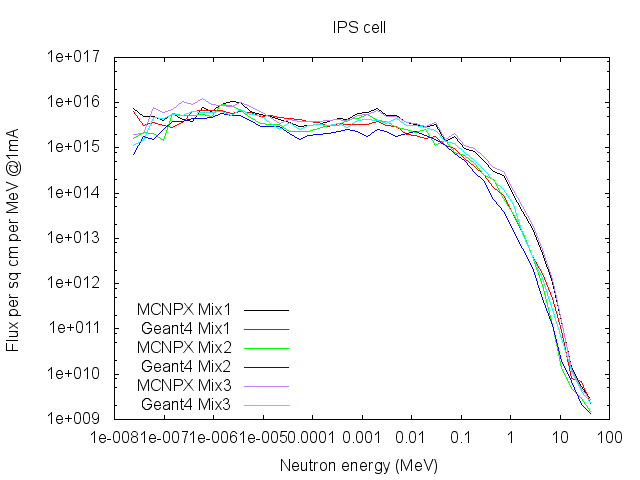}
\end{center}
\caption{Neutron flux averaged over the IPS cells }
\label{Fig:IPSFlux}
\end{figure}

  Spectra of IPS regions shown in Figure~\ref{Fig:IPSFlux} are taken as the average of 6 IPS cells spectra. Being `in pile' they have neutron fluxes similar to those in the fuel cells that surround them but with  fewer thermal neutrons. 
 
\begin{figure}[ht]
\begin{center}
\includegraphics[width=0.49\textwidth]{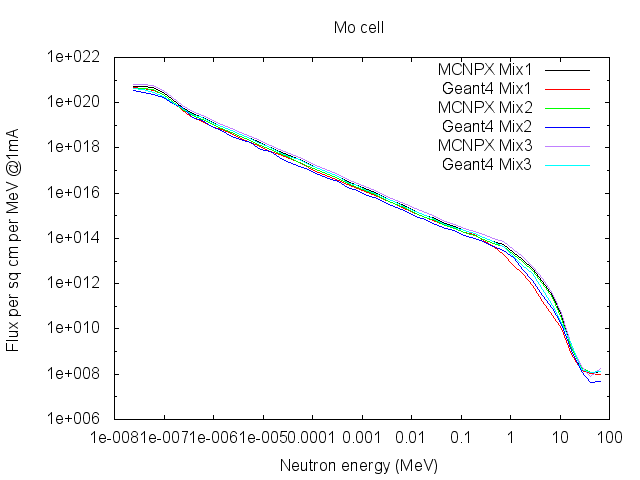}
\includegraphics[width=0.49\textwidth]{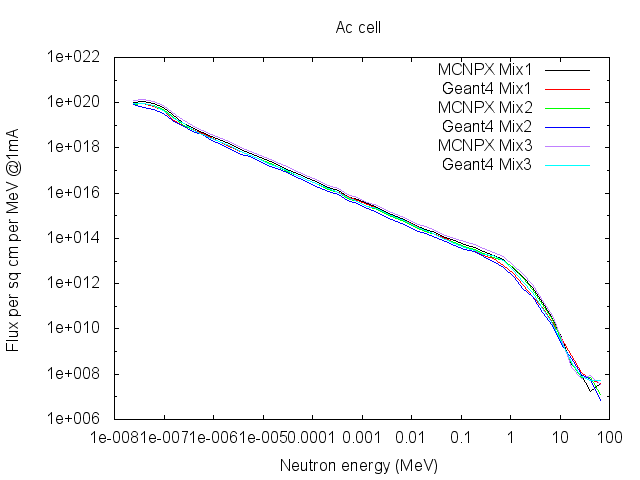}
\end{center}
\caption{Neutron flux averaged over the   Mo cells (left) and Ac cells (right)}
\label{Fig:MoAcFlux}
\end{figure}

The Mo and Ac production cells are farther away from the centre and the spectra, shown in Figure~\ref{Fig:MoAcFlux} in these regions are found to be softer, which is understandable as the neutrons reaching the outer cells have travelled further and undergone more collisions. 

Energy distribution of the neutron flux can be divided into the following three regions \cite{molina2017energy}:
\begin{enumerate}
  \item Thermal region - Neutron energy: $5\times 10^{-11}$ MeV to $5\times 10^{-7}$ MeV.
  \item Epithermal region - Neutron energy: $5\times 10^{-7}$ MeV to 0.5 MeV.
  \item Fast region - Neutron energy: 0.5 MeV to 20 MeV.
\end{enumerate}

Based on this, the flux percentage for each considered region with respect to the integral flux and the ratio between the fast and the thermal components are presented in Table~\ref{tab:FluxRegions}. The ratio depicts that the hardness of the spectra is 1\% to 2\% in IPS cells, 0.1\% for fuel cells and much smaller in Mo Ac cells.

\begin{table}[H]
\begin{center}
\begin{adjustbox}{width=0.85\textwidth}
\small
\begin{tabular}{|c|c|c|c|c|c|c|}
\hline
\multirow{2}{1cm}{Cell} 
 & \multirow{2}{1cm}{Fuel} 
  & \multirow{2}{1cm}{Program} & \multicolumn{3}{c|}{Percentage} & Ratio of fast\\
 \cline{4-6} 
  & & & Thermal & Epithermal & Fast &  to thermal \\
 \hline 
Fuel & Mix1 & {MX} & 63.36 & 36.60 & 0.04 & 0.0007 \\
 \cline{3-7} 
& & {G4} & 53.17 & 46.78  & 0.05 & 0.0010 \\
 \cline{2-7} 

&  Mix2  & {MX} & 61.25 & 38.70 & 0.04 & 0.0008 \\
 \cline{3-7} 
& & {G4} & 50.96 & 48.99 & 0.04 & 0.0008 \\
 \cline{2-7} 

&  Mix3 & {MX} & 82.89 & 17.09 & 0.03 & 0.0003 \\
 \cline{3-7} 
& & {G4} & 77.68 & 22.30 & 0.02 & 0.0002 \\
\hline

IPS & Mix1 & {MX} & 20.29 & 79.31 & 0.40 & 0.0195 \\
 \cline{3-7} 
& & {G4} & 21.48 & 78.33 & 0.20 & 0.0092 \\
 \cline{2-7} 

&  Mix2  & {MX} & 18.61 & 81.12 & 0.26 & 0.0142 \\
 \cline{3-7} 
& & {G4} & 20.27 & 79.58 & 0.14 & 0.0071 \\
 \cline{2-7} 

&  Mix3 & {MX} & 23.12 & 76.42 & 0.46 & 0.0197 \\
 \cline{3-7} 
& & {G4} & 21.23 & 78.48 & 0.29 & 0.0135 \\
\hline
Mo & Mix1 & {MX} & 96.98 & 3.02 & $8.142\times 10^{-6}$ & $8.396\times 10^{-8}$ \\
 \cline{3-7} 
& & {G4} & 97.50 & 2.50 & $4.333\times 10^{-6}$ & $4.443\times 10^{-8}$ \\
 \cline{2-7} 

&  Mix2  & {MX} & 96.99 & 3.01 & $8.297\times 10^{-6}$ & $8.555\times 10^{-8}$ \\
 \cline{3-7} 
& & {G4} & 96.80 & 3.20 & $7.417\times 10^{-6}$ & $7.662\times 10^{-8}$ \\
 \cline{2-7} 

&  Mix3 & {MX} & 96.98 & 3.02 & $8.721\times 10^{-6}$ & $8.992\times 10^{-8}$ \\
 \cline{3-7} 
& & {G4} & 95.97 & 4.03 & $9.506\times 10^{-6}$ & $9.905\times 10^{-8}$ \\
\hline
Ac & Mix1 & {MX} & 96.74 & 3.26 & $8.657\times 10^{-6}$ & $8.949\times 10^{-8}$ \\
 \cline{3-7} 
& & {G4} & 96.54 & 3.46 & $6.719\times 10^{-6}$ & $6.960\times 10^{-8}$ \\
 \cline{2-7} 

&  Mix2  & {MX} & 96.77 & 3.23 & $9.117\times 10^{-6}$ & $9.422\times 10^{-8}$ \\
 \cline{3-7} 
& & {G4} & 96.39 & 3.61 & $6.406\times 10^{-6}$ & $6.646\times 10^{-8}$ \\
 \cline{2-7} 

&  Mix3 & {MX} & 96.76 & 3.24 & $9.362\times 10^{-6}$ & $9.674\times 10^{-8}$ \\
 \cline{3-7} 
& & {G4} & 96.79 & 3.21 & $9.560\times 10^{-6}$ & $9.877 \times 10^{-8}$ \\
\hline
\end{tabular}
\end{adjustbox}
\caption{Neutron flux percentage for different regions using GEANT4 and MCNPX for different fuel Mix}
\label{tab:FluxRegions}
\end{center}
\end{table} 

\section{Results: Fuel evolution with thorium as fuel
}

\subsection{Method}
                                            
Fuel evolution takes place due to  fission,
$\alpha$ and $\beta$ decay,
  and reactions with neutrons.
So the number of atoms $X$ of some isotope in an element of a nuclear reactor changes
according to

\begin{equation}\label{eqn:DecayEq}
\frac{dX}{dt} = Q\textsubscript{1}Y +  \lambda\textsubscript{2} P - Q\textsubscript{3}X - \lambda\textsubscript{4} X
\end{equation}

where  $Y$ is the number of atoms that can produce $X$ by reacting with a neutron, and $P$ is the number of atoms decaying to $X$ at rate $\lambda_2$, while $X$ itself is absorbing neutrons (by (n,$\gamma$) or (n,2n) or other reactions)  at rate $Q_3$ and decaying at rate $\lambda_4$.  $Q$\textsubscript{1} is the creation  reaction 
probability $\int \sigma\textsubscript{1}  (E) \phi (E) dE $ and $Q$\textsubscript{3} is the  destruction probability  $\int\sigma\textsubscript{3}  (E) \phi(E) dE$, where $\sigma\textsubscript{1}$ and $\sigma\textsubscript{3}$ are the relevant cross sections and $\phi$ is the flux;  these terms involve different cross sections but the same neutron flux.

For a particular isotope not  all of the  terms  in equation~\ref{eqn:DecayEq} may be relevant, however these  four are the  maximum we need to consider (they can be extended in cases where $X$ is produced by more than one reaction, or as the product of more than one decay).  

\label{Algebraic}

Equation~\ref{eqn:DecayEq}  is a differential equation involving three unknowns: $X$, $Y$ and $P$. $Y$ and $P$ will have similar equations of their own,  and  these  may  involve  further  species.   To predict what will happen,  a complete  set  of equations needs to be written, one for each isotope involved, known as the Bateman Equations.

 Dropping the different names and calling the isotopes $X_1$ , $X_2$ ...$X_n$, or just  $\vec{X}$ there
are $n$ equations  for the $n$ isotopes.  The equations  have the simple matrix form: 

\begin{equation}
\frac{d\vec{X}}{dt} =  M \vec{X} 
\end{equation}

where $M$ is a matrix, of which  the  elements are decay rates $\lambda$ or reaction  probabilities $\int\sigma (E)\phi(E) dE $.  

 These equations can be solved either by stepwise integration (Euler's method) or by an exact algebraic technique:
 the eigenvectors $u_i$ and eigenvalues $a_i $ of $M$ are found (we use the  R package {\tt eigen}),
 the initial composition is expressed as a sum over $u_i$, $\vec X(0)=\sum_i C_i u_i$ and the mixture at time is then just $\vec X(t)=\sum_i C_i u_i e^{a_i t}$.
 
 In Equation~\ref{eqn:DecayEq} and its generalisations the lifetime $\lambda$ terms are taken from known data.
 The reaction $Q$ terms depend on the neutron flux, convoluted with the appropriate cross section.
 In MCNPX these are provided on demand using 'Tally cards'. For Geant4 they had to be constructed 
 by numerical integration of the flux obtained by the simulation program and the appropriate cross section as given by the JEFF3.1  library~\cite{JEFF}. 
 As these two sets of data did not generally use the same energy bins, interpolation was done using using the R function {\tt approx}.
 
 This solution uses the flux as determined above in section~\ref{Spectra}. Changes in the composition will lead to changes in the flux, so these calculations are not exact; nevertheless the changes are small and will not have a large  effect on the results, so the approximation is accurate enough to be useful. 

When the  evolution $\vec{X}$(t) is plotted then  typically  
one sees the  $e ^ {-\lambda  t}$   decay  of the components  one starts with,  and  the  rise and  then  fall , $ e ^ {-\lambda t}  (1-e ^ {-\lambda ' t}) $ of the  daughter products.

\subsection{Fertile to Fissile conversion}

Turning to the specific cases considered here, these involve the conversion of fertile isotopes to the intermediate nuclei through neutron capture events. For mix 1, this is the conversion of \Iso 238 U \ to \Iso 239 U \ while 
for mix 2 and mix 3 it is \Iso 232 Th \ to \Iso 233 Th . Rapid beta decay in both these cases leads to the formation of \Iso 239 Np \ and \Iso 233 Pa \ respectively. The half-life of \Iso 239 Np \ is 23 minutes and that of \Iso 233 Pa \ is 21 minutes, so these can be considered as instantaneous.
The second beta decay transforms \Iso 239 Np \  to \Iso 239 Pu , which is still rapid with a half-life of 2.4 days, whereas for mix 2 and mix 3, conversion of \Iso 233 Pa \ to \Iso 233 U \ is relatively slower with a half-life of 27 days. Further, if \Iso 233 Pa \ absorbs a neutron in an intense neutron flux, it could lead to the process diverting in unwanted directions. This is an issue well known for thorium reactors \cite{barlow2016studies}.

\begin{figure}[ht]
\begin{center}
\includegraphics[width=0.6\textwidth,clip,trim={0 0 100 0}]{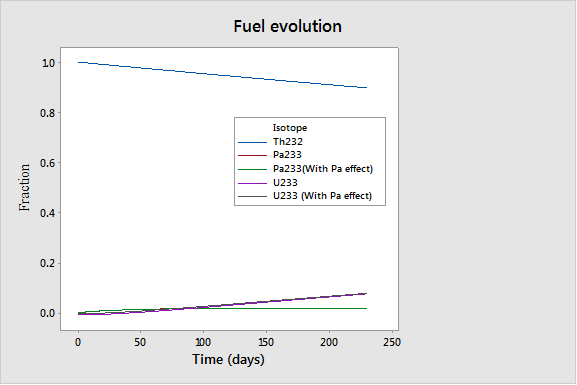}
\end{center}
\caption{Thorium fuel evolution with and without \Iso 233 Pa \ neutron absorption effect}
\label{Fig:PaEffect}
\end{figure}

GEANT4 and MCNPX calculated neutron flux convolved with the cross section (as per equation~\ref{eqn:WeightedFlux}) for the three relevant isotopes are presented in Table~\ref{tab:nAbsRate}. The two programs GEANT4 and MCNPX agree that the conversion rates differ for the three fuel mixtures. As the composition changes, the probability of fertile to fissile thorium conversion increases. The protactinium absorption probabilities (indicating the loss of \Iso 233 Pa \ from the chain) are relatively large compred to the probabilities of thorium nuclei entering the chain, but the  amount of \Iso 233 Pa \ in the fuel at any time is very small. The \Iso 233 Pa \ effect (neutron absorption in the intermediate state) can be evaluated by solving the Bateman equation over 230 days for 2.5 mA beam current. This is plotted in Figure~\ref{Fig:PaEffect}. The  effect is observed to be very small as \Iso 233 Pa \ stabilizes soon.

\begin{table}[ht]
\begin{center}
\begin{tabular}{|c|c|c|c|c|c|c|c|}
\hline
\multirow{2}{2cm}{Fuel} & \multicolumn{2}{c|}{U238} & \multicolumn{2}{c|}{Th232} & \multicolumn{2}{c|}{Pa233} \\
 \cline{2-7} 
& G4 & MX & G4 & MX & G4 & MX \\
 \hline 
U/Pu (Mix1) & 0.026 & 0.028 & 0.034 & 0.045 & 0.075 & 0.18\\
 \hline
Th/Pu (Mix2) & 0.032 & 0.036 & 0.028 & 0.028 & 0.061 & 0.15\\
 \hline
Th/U (Mix3) & 0.035 & 0.046 & 0.035 & 0.039 & 0.076 & 0.19 \\
 \hline
\end{tabular}
\caption{Neutron absorption rates for isotopes in the fuel cells for 1 mA beam}
\label{tab:nAbsRate}
\end{center}
\end{table}

\subsection{Production of \texorpdfstring {\Iso 232 U} {232-U} }

The isotope \Iso 232 U \ is important 
in the thorium cycle because it is inevitably co-produced with \Iso 233 U, and  decays with the relatively short half live of 69 years, with a long decay chain involving high energy gamma rays  \cite{Kang} which can make handling in conventional glove boxes impossible, and can also damage electronics.  This can be viewed as an advantage in that it is makes weaponising fissile \Iso 233 U \ technically very difficult. It is, however, a possible problem in a thorium fuelled reactor.

There are 3 routes for its production, starting from isotopes already mentioned:

\begin{enumerate}
    \item \Iso 233 U \ 
     $\begin{matrix}n,2n\\ \longrightarrow \\
     \\
    \end{matrix}$ \  
    \Iso 232 U
    
    \item
    \Iso 233 Pa \   $\begin{matrix}n,2n\\ \longrightarrow \\
     \\
    \end{matrix}$ \  
    \Iso 232 Pa \ 
    $\begin{matrix}\beta^-\\ \longrightarrow \\
     \\
    \end{matrix}$ \ \Iso 232 U

\item \Iso 232 Th \ 
$\begin{matrix}n,2n\\ \longrightarrow \\
     \\
    \end{matrix}$ \  
    \Iso 231 Th  \ 
    $\begin{matrix}\beta^-\\ \longrightarrow \\
     \\
    \end{matrix}$ \ \Iso 231 Pa 
     \ 
    $\begin{matrix}n,\gamma\\ \longrightarrow \\
     \\
    \end{matrix}$ \ \Iso 232 Pa \ 
     \ 
    $\begin{matrix}\beta^-\\ \longrightarrow \\
     \\
    \end{matrix}$ \ \Iso 232 U
    
\end{enumerate}

\begin{table} [ht]
\begin{tabular}{|c|c|c|c|c|}
\hline
Reaction & \Iso 233 U (n,2n) \Iso 232 U & \Iso 233 Pa (n,2n) \Iso 232 Pa 
& \Iso 232 Th (n,2n) \Iso 231 Th &
\Iso 231 Pa (n,$\gamma$) \Iso 232 Pa \\ 
\hline
MCNPX     &  $4.58 \times 10^{-5}$ & $1.74 \times 10^{-4}$ & $2.27 \times 10^{-4}$ &
$1.41 \times 10^{-1}$
\\
GEANT4     & $1.70 \times 10^{-4}$ & $4.33 \times 10^{-4}$ &  $5.65 \times 10^{-4}$&  $1.44 \times 10^{-1}$\\
\hline
\end{tabular}
\caption{\label{tab:U232} Production probabilities for \Iso 233 U .The numbers in the table are as given by the MCNP tally, namely the flux in ${\rm cm}^{-2}$ multiplied by the cross section in barns. } 
\end{table}

The relevant production probabilities are shown in Table~\ref{tab:U232}.
The rate for the \Iso 231 Pa \ absorption is in good agreement between the two simulations (the MCNPX numbers quote a typical error of 1\%) but for the $(n,2n)$ probabilities there are differences. This was traced to the fact that these cross sections only become significant at 
high neutron energies ($\sim 10$ MeV), at which point, as can be seeing from figure~\ref{Fig:Fuel}, the flux is falling rapidly. MCNPX uses the `tally' feature to record each individual track, weighted by the cross section whereas for Geant4 the 
neutron spectrum was histogrammed, and the production probabilities obtained by convoluting this with the tabulated energy-dependent cross section (taken from the ENDF70 library~\cite{endf70}). These tables are fine-grained at low energies with 
only a few values tabulated at high energies. 
For typical $(n,\gamma)$ reactions this is adequate, but 
for these $(n,2n)$ reactions 
the bin width is not narrow enough to describe the necessary detail.   Hence the MCNPX numbers are used in the rest of this section.

\begin{figure}[ht]
    \centering
    \includegraphics[width=10 cm]{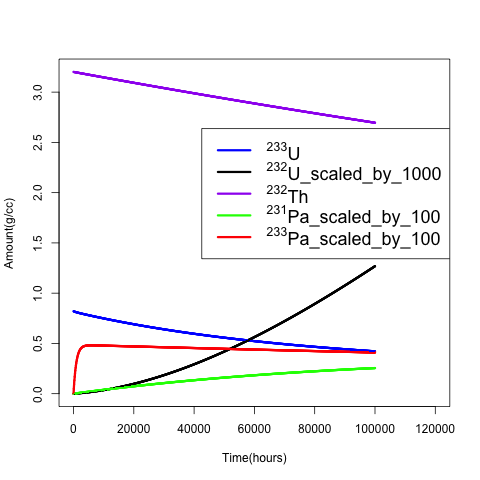}
    \caption{Evolution of a fuel rod, showing growth of \Iso 232 U }
    \label{fig:Bateman5}
\end{figure}

The values in the tables are multiplied by $10^{-24}$ to be converted into a dimensionless number: the probability that a particular target nucleus will be converted in the specified way by a single beam proton.
To find the total rate one multiplies by the density for a nominal 2.5 mA beam.
These were put into the Bateman equations and the resulting time evolution is shown in figure~\ref{fig:Bateman5}

 This considers fuel element evolving over a long period without replacement (100,000 hours is over 11 years).
 The intermediate \Iso 232 Pa \ and \Iso Th 231 \ decays, of the order of 1 day, are treated as instantaneous.
 It does not take into account changes in the flux due to fuel composition, or the accumulation of fission products, so it should be taken as indicative. 
 
 The largest contribution to \Iso 233 U \ growth comes from the \Iso 231 Pa . There is some contribution from the \Iso 233 U \ source, but the \Iso 233 Pa \ contribution is negligible.
 
A density of 1 mg/cc of \Iso 233 U , reached by the fuel after $\sim 10$ years, corresponds to an activity of 1.19 GBq/cc, which is well above the threshold for Intermediate Level Waste of 12 GBq/tonne. It will require careful handling and disposal.

\section {Results: Incineration of Minor Actinides}

\begin{figure}[ht]
\begin{center}
\includegraphics[width=0.4\textwidth]{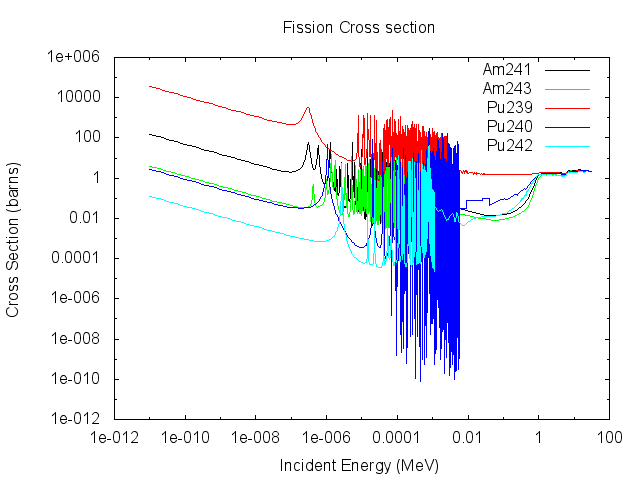}\ 
\includegraphics[width=0.4\textwidth]{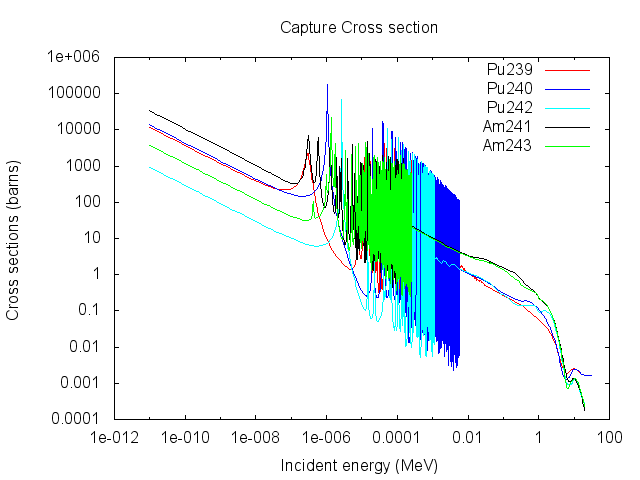}
\end{center}
\caption{Fission and Absorption cross sections as functions of neutron energy, for various minor actinides, using the JEFF 3.1 library\cite{JEFF}}
\label{Fig:Fission}
\end{figure}

From the neutron flux calculated in Section~\ref{Spectra} the actinide burning rates in the reactor can be calculated. The absorption and fission cross sections, taken from the JEFF 3.1 library \cite{JEFF,JEFFGeant} for different minor actinides are illustrated in Figure~\ref{Fig:Fission} . 

From these figures it can be seen that the fission cross section dominates over the absorption cross section above 1 MeV, demonstrating  the advantage of fast neutrons for incineration. Such a hard neutron spectrum was observed in the fuel cell and IPS cells, in Figures~\ref{Fig:Fuel} and  \ref{Fig:IPSFlux}.  The spectra in the Mo and  Ac cells in Figure~\ref{Fig:MoAcFlux} are not as hard as in the IPS and fuel cells but since these cells are available for irradiation, we also consider these locations for burn up studies. 

The weighted flux $F$ in a volume $V$ is given by the convolution with the cross section
\begin{equation}\label{eqn:WeightedFlux}
F = \dfrac{1}{V}\int_0^{E\textsubscript{max}}{\phi(E)\sigma(E)dE}
\end{equation} 

where $V$ is in cm\textsuperscript{3}, $\phi(E)$ is the total track length of the neutrons produced by a single beam particle, so $\phi(E)/V$ is the flux and $\sigma$(E) is the reaction cross section (which needs to be multiplied by a factor 10$^{-24}$  to convert barn to cm\textsuperscript{2}).
$F$ is thus  
\cite{barlow2016studies}
  the probability that a given MA nucleus will be transformed in the neutrons produced by one beam particle. When multiplied by the beam current (in protons per second) this gives the burn-up rate: the inverse of the mean life time of the MA in the reactor.

This weighted flux is calculated by GEANT4 and MCNPX  for a number of significant actinide species: \Iso 241 Am , \Iso 243 Am , \Iso 237 Np , \Iso 239 Pu , \Iso 240 Pu \ and \Iso 242 Pu \ for both the inner IPS cells and the outer cells used for Mo and Ac production (the two types are so similar they were combined). Three fuel mixtures described earlier are considered for these calculations. 
Results are shown in Table~\ref{tab:nAbsRateInner} and Table~\ref{tab:nAbsRateOuter} respectively. The ratio in 7\textsuperscript{th} and 8\textsuperscript{th} column corresponds to the ratio of fission to absorption, according to the two programs. It tells whether fission or absorption is dominating. 

The tables show that
\begin{enumerate}
\item Results of the two programs agree reasonably well: the difference shows the uncertainties inherent in such calculations. 
  \item Slightly higher burn up rates are predicted for Mix 3, according to MCNPX.
  \item \Iso 239 Pu \ being important as a fuel rather than a waste product, is a special case. Its fission probability is higher than absorption, in the inner cells where the spectra are harder, as well as in the outer cells.
  \item For the other isotopes in the table, the ratio of fission to absorption is respectable in the IPS cells but it is negligible in the outer cells.   Exposing such isotopes in the inner IPS cell will either transmute them to short lived fission products and/or to some higher atomic weight nuclei through neutron absorption.   The ratio in outer cells suggest limited use of these cells for incineration.
\end{enumerate}
\begin{table}[ht]
\begin{center}
\begin{adjustbox}{width=1\textwidth}
\small
\begin{tabular}{|c|c|c|c|c|c|c|c|}
\hline
\multirow{2}{1cm}{Isotope} 
 & \multirow{2}{1cm}{Fuel} & \multicolumn{6}{c|}{Inner IPS} \\
 \cline{3-8} 
  & & Fission MX & Fission G4 & Absorption MX & Absorption G4 & Ratio MX & Ratio G4 \\
 \hline 
\Iso 239 Pu  & U/Pu & 0.23 & 0.12 & 0.04 & 0.02 & 5.57 & 4.92 \\ 
&  Th/Pu & 0.19 & 0.24 & 0.03 & 0.04 & 5.82 & 5.64 \\
&  Th/U & 0.27 & 0.16 & 0.05 & 0.03 & 6.09 & 4.73 \\
\hline
\Iso 240 Pu  & U/Pu & 0.06 & 0.02 & 0.04 & 0.02 & 1.49 & 1.00 \\
&  Th/Pu & 0.06 & 0.07 & 0.03 & 0.05 & 1.62 & 1.43 \\
&  Th/U & 0.06 & 0.04 & 0.05 & 0.03 & 1.15 & 1.19 \\
\hline
\Iso 242 Pu  & U/Pu & 0.05 & 0.02 & 0.05 & 0.03 & 	0.98 & 0.62 \\			&  Th/Pu & 0.04 & 0.05 & 0.04 & 0.05 & 1.02 & 1.02 \\						&  Th/U & 0.06 & 0.02 & 0.06 & 0.04 & 1.15 & 0.59 \\
\hline
\Iso 241 Am  & U/Pu & 0.05 & 0.02 & 0.21 & 0.12 & 0.24 & 0.15 \\	
&  Th/Pu & 0.04 & 0.05 & 0.16 & 0.19 & 0.26 & 0.23 \\
&  Th/U & 0.06 & 0.02 & 0.23 & 0.16 & 0.28 & 0.14 \\
\hline
\Iso 243 Am  & U/Pu & 0.04 & 0.01 & 0.20 & 0.12 & 0.18 & 0.11 \\
&  Th/Pu & 0.03 & 0.03 & 0.15 & 0.18 & 0.20 & 0.17 \\
&  Th/U & 0.05 & 0.02 & 0.21 & 0.15 & 0.21 & 0.10 \\
\hline	
\end{tabular}
\end{adjustbox}
\caption{Neutron absorption rates for isotopes in the inner IPS cells using GEANT4 and MCNPX for 1 mA beam. Units are probability per beam proton times $10^{24}$.}
\label{tab:nAbsRateInner}
\end{center}
\end{table} 
\begin{table}[ht]
\begin{center}
\begin{adjustbox}{width=1\textwidth}
\small
\begin{tabular}{|c|c|c|c|c|c|c|c|}
\hline
\multirow{2}{1cm}{Isotope} 
 & \multirow{2}{1cm}{Fuel} & \multicolumn{6}{c|}{Outer Mo Ac cells} \\
 \cline{3-8} 
  & & Fission MX & Fission G4 & Absorption MX & Absorption G4 & Ratio MX & Ratio G4 \\
 \hline 
\Iso 239 Pu  & U/Pu & 5.78 & 4.43 & 2.91 & 2.22 & 1.98 & 2.00 \\
&  Th/Pu & 4.70 & 3.51 & 2.37 & 1.89 & 1.99 & 1.85 \\
&  Th/U & 6.50 & 4.10 & 3.37 & 2.15 & 1.93 & 1.91 \\
\hline
\Iso 240 Pu  & U/Pu & 0.01 & 0.00 & 3.80 & 2.77 & 0.00 & 0.00 \\
&  Th/Pu & 0.01 & 0.01 & 3.08 & 2.36 & 0.00 & 0.00 \\
&  Th/U & 0.01 & 0.01 & 4.67 & 3.22 & 0.00 & 0.00 \\
\hline
\Iso 242 Pu  & U/Pu & 0.01 & 0.00 & 0.26 & 0.19 & 0.03 & 0.01 \\
&  Th/Pu & 0.01 & 0.00 & 0.21 & 0.15 & 0.03 & 0.02 \\	
&  Th/U & 0.01 & 0.01 & 0.32 & 0.23 & 0.03 & 0.02 \\
\hline
\Iso 241 Am  & U/Pu & 0.06 & 0.04 & 7.86 & 5.77 & 0.01 & 0.01 \\	
&  Th/Pu & 0.05 & 0.04 & 6.36 & 5.43 &  0.01 & 0.01 \\
&  Th/U & 0.07 & 0.05 & 9.18 & 6.57 & 0.01 & 0.01 \\
\hline
\Iso 243 Am  & U/Pu & 0.01 & 0.00 & 2.04 & 1.50 & 0.00 & 0.00 \\
&  Th/Pu & 0.01 & 0.00 & 1.66 & 1.26 & 0.00 & 0.00 \\
&  Th/U & 0.01 & 0.01 & 2.51 & 1.86 & 0.00 & 0.00 \\
\hline	
\end{tabular}
\end{adjustbox}
\caption{Neutron absorption rates for isotopes in the outer Mo Ac cells using Geant4 and MCNPX for 1 mA beam. Units are probability per beam proton times $10^{24}$.}
\label{tab:nAbsRateOuter}
\end{center}
\end{table}
\subsection{Americium incineration}
\label{Am241Incineartion}
\begin{figure}[ht]
\begin{center}
\includegraphics[height=.35\linewidth]{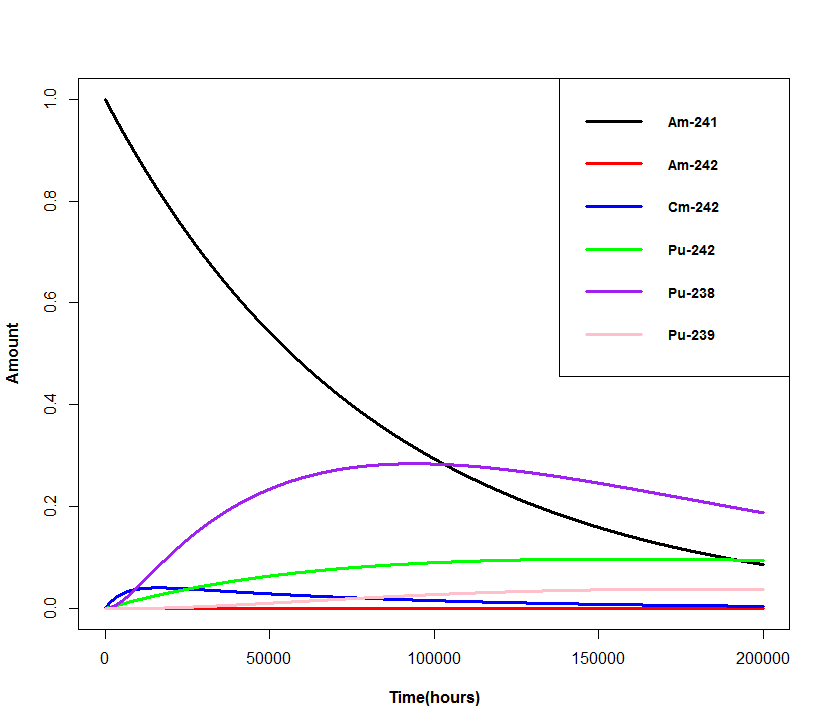}
\includegraphics[height=.35\linewidth]{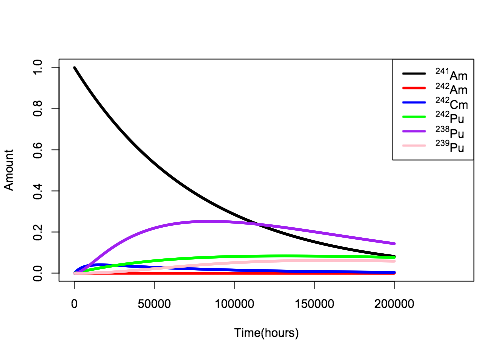}
\end{center}
\caption{Evolution of \Iso 241 Am \ and its products: GEANT4 (Left) and MCNPX  (Right)}
\label{Fig:Am241}
\end{figure} 
\begin{figure}[ht]
\begin{center}

\includegraphics[width=.6\linewidth]{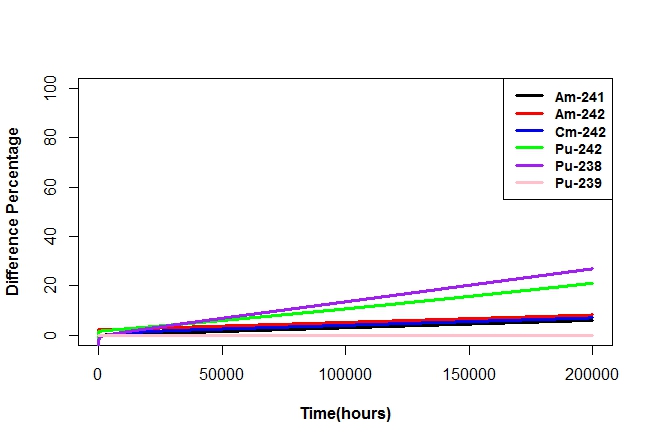}
\end{center}
\caption{Evolution of \Iso 241 Am \ and its products: percent difference between GEANT4 and MCNPX  }
\label{Fig:Am241diff}
\end{figure} 
The incineration of \Iso 241 Am \ in the IPS cell with Mix2 using GEANT4 and MCNPX, based on these burn up rates, is shown in Figure~\ref{Fig:Am241}.  The deviations, shown in Figure~\ref{Fig:Am241diff}, are typically a few percent.  The Bateman equations were solved over the period of 23 years by eigenvalue method described in section \ref{Algebraic}. The reaction rates, the $Q$ factors in equation~\ref{eqn:DecayEq}, are obtained from the tables and the decay rates are the inverse of the lifetimes of the isotopes. Single absorption and subsequent decays are considered. As discussed in subsection \ref{MATrans}, \Iso 241 Am \ is consumed by absorption rather than fission. \Iso 238 Pu \ is the main product in this process which is produced from the successive decays of \Iso 242 Am \ and \Iso 242 Cm . Other isotopes are also produced in smaller quantities. 

These solutions are indicative in that they assume continuous operation of the reactor, and they ignore changes in the composition of the fuel (including the accumulation of fission products). To include such efforts would require many assumptions about the duty cycle and fuel processing strategy, and would not effect the overall conclusions. This shows that the typical minor actinide \Iso 241 Am \ would be incinerated in MYRRHA. There would be small quantities of other isotopes produced but they present less of a problem as regards storage. However the process would be slow. An economically effective \Iso 241 Am \ burner would need to operate at much higher currents. However, the MYRRHA reactor would be a useful prototype.

\subsection{Minor actinide production and net incineration}

\begin{figure}[ht]
\begin{center}
\includegraphics[height=.42\linewidth]{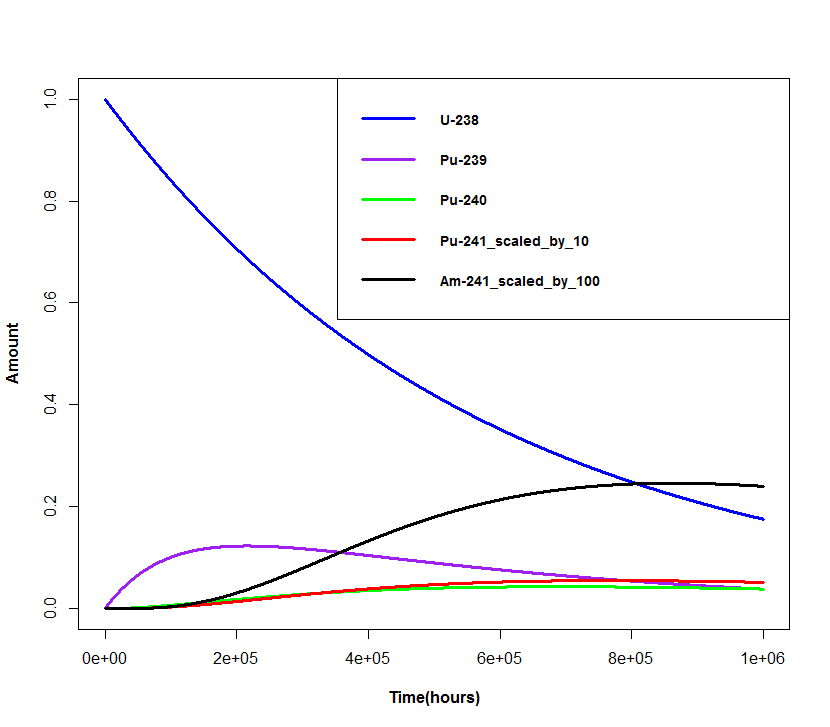}\includegraphics[height=.42\linewidth]{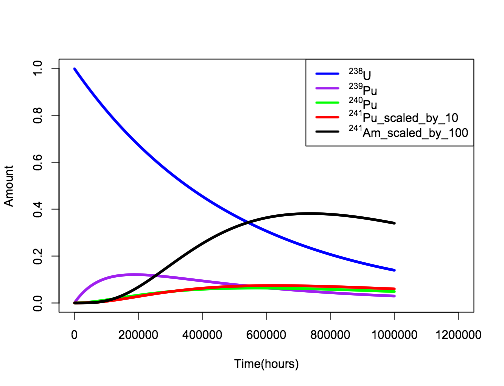}
\end{center}
\caption{Evolution for \Iso 238 U \ and its products, Geant4 (Left) and MCNPX (Right) }
\label{Fig:UraniumCycle}
\end{figure}
\begin{figure}[ht]
\begin{center}
\includegraphics[height=0.4\linewidth]{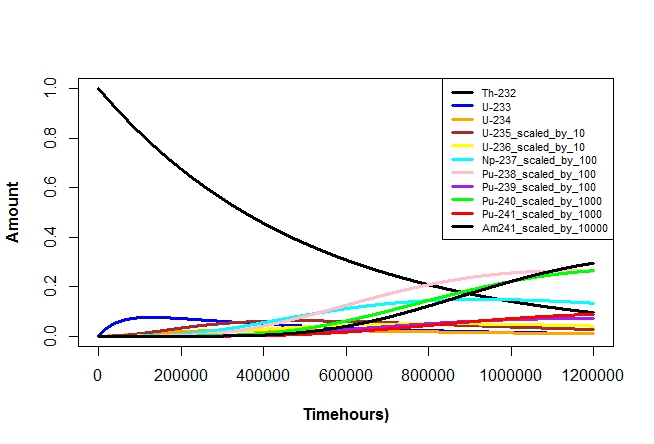}\includegraphics[height=0.4\linewidth]{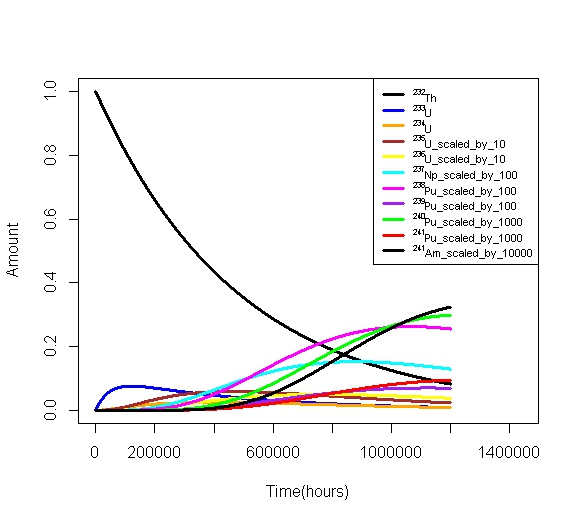}
\end{center}
\caption{Evolution for \Iso 232 Th \ and its products, GEANT4 (Left) and MCNPX (Right) }
\label{Fig:ThCycle}
\end{figure} 
In order to compare the incineration capability of uranium and thorium based fuel, it is important to investigate the net incineration in the two fuel cycles. The amount of minor actinide produced is compared with that incinerated, as calculated in subsection \ref{Am241Incineartion}. In the same model the formation of \Iso 241 Am \ starting from uranium or thorium is considered, using GEANT4 and MCNPX. The assumptions made in this modelling are: 

\begin{enumerate}
 \item Short life time decays are considered spontaneous.
 \item Accelerator operation at 2.5 mA is continuous.
  \item The entire fuel is treated as a single lumped object.
  \item  Variation in the neutron spectrum during fuel evolution is neglected.
  \item A nominal timescale of 100 years.
\end{enumerate}

Results of GEANT4 and MCNPX for uranium and thorium fuel cycles are shown in Figure~\ref{Fig:UraniumCycle} and \ref{Fig:ThCycle}. From the figures, it is clear that both the codes agree well for the trends in the two fuel cycles. It can be noted from Figure~\ref{Fig:UraniumCycle} that the amount 
of \Iso 241 Am \ produced from a Uranium base needs to be scaled by factor of $\sim 100 $ to be visible. So $\sim 0.4 $ \% of \Iso 238 U \ transforms into \Iso 241 Am \ and although this is a small fraction, given the large amount of \Iso 238 U \ in the fuel, it is producing a significant amount of \Iso 241 Am . On the other hand, \Iso 241 Am \  production from thorium (Figure~\ref{Fig:ThCycle}) needs scaling by a factor of 10000 to be visible. Minor actinide production is decreased by a factor of $\sim 100 $ when using thorium based in place of uranium based fuels.

\section{Conclusions and future work}
\begin{enumerate}
\item The reactor geometry has been successfully implemented  using GEANT4.
 \item The neutron flux spectra obtained by different simulation programs GEANT4 and MCNPX are benchmarked against each other and show good agreement for neutron flux spectra at various location of the reactor. The fuel spectra by these two programs also agree with other studies (\cite{sarotto2013myrrha}).
  
 \item However there are some differences in the predictions, for some quantities in some regions, showing the usefulness of having two simulation programs to point out where predictions can be less reliable.

  \item Broad features of neutron spectra are same for the three fuel mixes at all the chosen locations of the reactor. This indicates that provided the reactivity remains the same (in this case $K$\textsubscript {eff} = 0.95), the spectra is insensitive to different fuel composition. Nevertheless they differ in detail which could be due to different absorption cross sections of the isotopes.
   \item GEANT4 and MCNPX results presented in Table~\ref{tab:nAbsRateInner} and Table~\ref{tab:nAbsRateOuter} depict that MYRRHA can convert measurable amount of minor actinide waste hereby proving the concept of industrial transmutation system.
  \item Net incineration in case of \Iso 241 Am \ is improved with thorium fuels as the amount of \Iso 241 Am \ produced is much smaller than the amount incinerated.
  \item The spectrum is harder and greater near the centre of the reactor. While the fuel cells are full (with fuel) there is space in the IPS that can be used for incineration studies.
 \item The slow timescale in the study shows that long term waste problem is not going to be solved by MYRRHA itself, however it can be utilized as a demonstrator and a prototype for large scale systems that operate at much higher currents.

\end{enumerate}
It is suggested to look at the fission products and burn up using the codes such as FISPACT or ALEPH to make time dependence more meaningful. But it will require detailed assumptions about the operational cycles and refuelling which were beyond the scope of this work.
\section*{Acknowledgements}

The authors would like to thank 
the staff of MYRRHA, espeically Hamid A\" it Abderrahim, Edouard Malumbo and Gert van den Eynde, for providing data and for their comments and support.

\bibliographystyle{unsrt}
\bibliography{references} 
\end{document}